\documentclass[aip,graphicx]{revtex4-2}
\usepackage{graphicx}
\usepackage{dcolumn}
\usepackage{bm}
\usepackage{hyperref}
\draft 

\begin{document}
	
	
	\title{FPGA-based residual amplitude modulation suppression and control for compact atomic clocks} 
	
	
	
	\author{Tin Nghia Nguyen}
	\affiliation{Department of Physics, University of Colorado, Boulder, Colorado 80309-0390, USA}
	
	\author{Thomas R. Schibli}
	\email{trs@colorado.edu}
	\affiliation{Department of Physics, University of Colorado, Boulder, Colorado 80309-0390, USA}
	\affiliation{JILA, NIST, and the University of Colorado, Boulder, Colorado, 80309-0440, USA}
	
	\date{\today}
	
	\begin{abstract}
		We designed an FPGA fabric to provide phase modulation techniques to lock lasers to optical frequency references. The method incorporates an active residual-amplitude-modulation (RAM) suppression scheme that relies on complex modulation. All the required servos to construct an optical atomic clock are incorporated onto the same low-cost, commercial FPGA chip. We demonstrate a reliable, long-term RAM suppression 60~dB with the remaining RAM level at -100~dBc and an improved stability of three decades when applied on a two-photon rubidium clock.
	\end{abstract}
	
	
	\maketitle 
	
	\section{Introduction}
	Compact atomic clocks, with sizes typically measured in liters, interrogate optical transitions of warm atomic (or molecular) vapors and deliver fractional instabilities at the order of 10$^{-12}$ to 10$^{-15}$.\cite{Phelps2018,Martin2018,JOKARUS2019} Such clocks rely on frequency and phase modulation techniques to obtain shot-noise limited sensitivities that are several orders of magnitude higher than traditional laser absorption approaches.\cite{Gehrtz1985} However, frequency and phase modulation schemes typically suffer from residual amplitude modulation (RAM), which occurs when the modulation sidebands are not both exactly equal in magnitude and opposite in phase.\cite{Zhang2014} This unwanted residual amplitude modulation causes a time-varying offset to the servo error signal used for stabilizing the laser frequency. This leads to drifts in the laser lock-point relative to the center frequency of the optical transition, and thus, degrades the clock’s stability.
	
	Several methods have been demonstrated to passively reduce or actively cancel the amount of RAM caused by electro-optic (EO) phase modulators. These include, but are not limited to, employing EO crystals that have lower temperature-induced birefringence,\cite{Jin2021} cutting crystals to have wedged input and output surfaces to remove etalon effects,\cite{Bi2019} and most commonly applying a bias DC-voltage across the EO crystal together with stabilizing its temperature.\cite{Zhang2014, Gillot2022} While the aforementioned methods work well, the first two methods employed free-space EO phase modulators that are much bigger and more prone to misalignment than fiber-based modulators. Besides, the last method requires a strict control of the modulator's temperature to prevent the feedback bias DC-voltage from going off limits and causing jumps in RAM level, which limits its applications outside temperature-controlled environment.
	
	Recently, another promising method of suppressing RAM by using complex modulation was proposed and demonstrated.\cite{Aronson2019, Chia2022} In this method, an EO amplitude modulator was used to apply intensity modulation that is exactly equal in magnitude but opposite in phase to the RAM created by an EO phase modulator to completely cancel the phase modulator’s parasitic RAM. This would suppress the RAM created by an EO phase modulator of any design at the cost of the additional optical insertion loss of the additional amplitude modulator. Despite its usefulness, this method was, to our knowledge, never shown to improve stability of any laser locking system, whether the laser being locked to an ultra-stable cavity or a narrow atomic/molecular transition. Furthermore, this method relies on a rather involved setup of a lock-in amplifier to demodulate the RAM signal and numerous synchronized function generators to create the RAM cancellation signal, all of which would significantly add to the size of a compact atomic clock, if realized with off-the-shelf components. If realized purely in analogue circuitry, the stability of phase adjustments and the flicker noise present in doubly-balanced microwave mixers can severely impact the overall performance of this approach. Here, we demonstrate that the entire complex modulation scheme for RAM suppression together with all of the servos and error signal generation blocks to lock a laser to an atomic transition can be fitted inside a low-cost field programmable gate array chip, which is, besides the flicker noise of the analog to digital converters (ADC) and the digital to analog converters (DAC) needed to interface to the experimental setup, free of drifts and flicker noise. Below we demonstrate the effectiveness of this method to improve the stability of a two-photon rubidium clock.
	
	\section{The FPGA fabric}
	The FPGA board in use is a commercial off-the-shelf (COTS) board from Koheron (model ALPHA250) that offers two 14-bit RF-ADCs and two 16-bit RF-DACs, each running at 250~MSpS. The ADCs are connected to the FPGA fabric via a randomized 14-bit differential DDR interface to provide very low digital noise at low latencies, and the DACs are connected via a single-ended 32-bit parallel interface for minimal latency, which results in an analog loopback latency of approximately 90~ns. These boards offer an additional four slow precision 24-bit ADCs and four, 244~kSpS precision 16-bit DACs, which can for instance be used for long-term power monitoring and beam pointing corrections in the clock setup. The precision DACs are sufficiently fast to be used to lock the frequency of a fiber laser to the atomic clock transition (see below). It is worth noting that the achievable loop bandwidth in such clocks is typically limited by the signal-to-noise ratio (SNR) from the spectroscopic signal. In traditional fluorescence monitoring Rb-vapor clocks, the SNR typically limits the achievable loop bandwidth to a few 10~kHz and hence, the 244~kSpS precision DACs provide sufficient bandwidth to lock the clock laser to the atomic transition. Using the slower DACs frees up the required RF resources to perform the modulation and RAM suppression described in the next paragraphs. All the analog periphery is connected to a Xilinx Zynq 7020-2 system on a chip (SoC) consisting of an Artix-7 FPGA and a dual-core ARM processor (Cortex-A9) running Linux with a custom TCP/IP socket for remote monitoring and control via a PC-based user interface written in Qt/C++.
	
	\begin{figure}
		\centering\includegraphics[width = 0.8\textwidth]{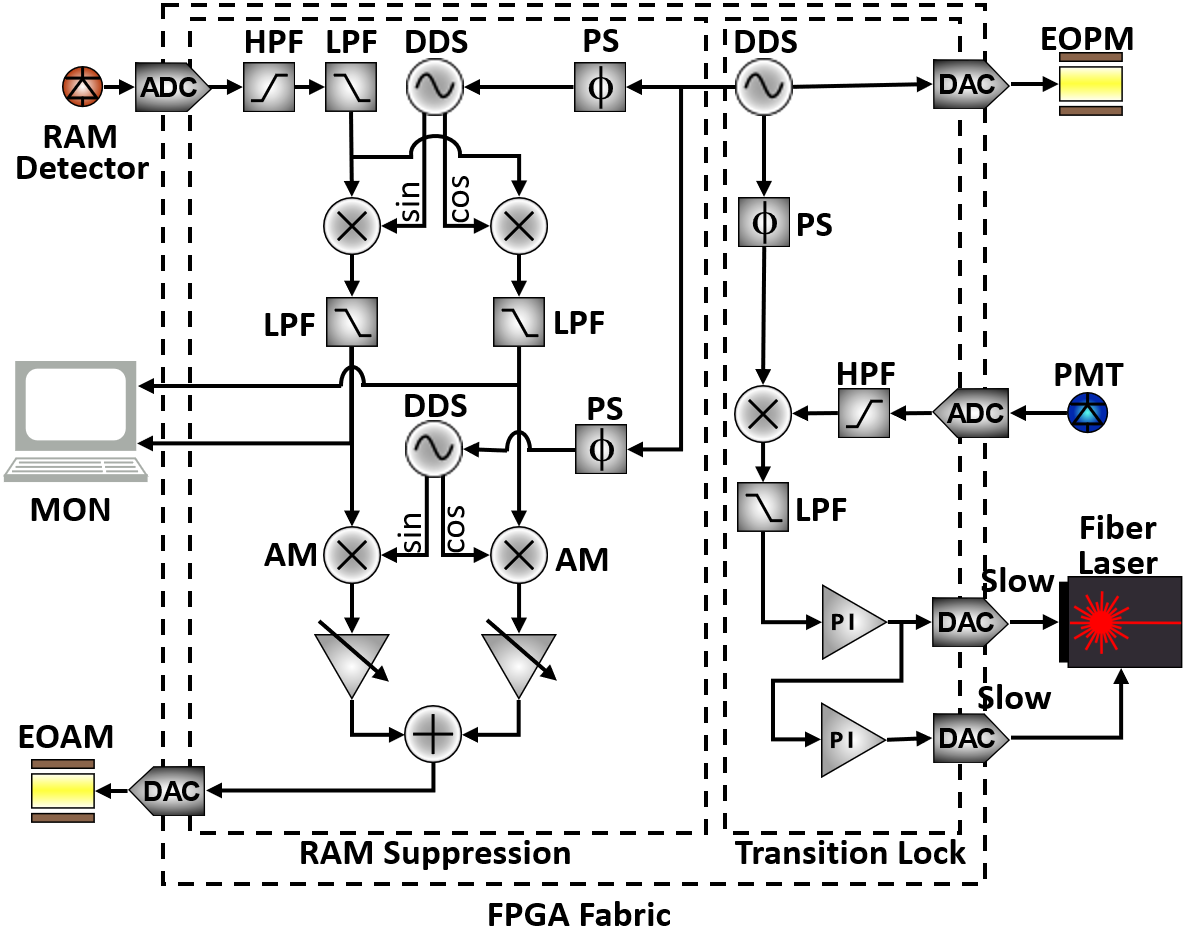}
		\caption{\label{fig:FPGA_Fabric} FPGA circuits for generating error signal, locking the laser to the two-photon rubidium transition, and suppressing RAM. ADC: analog to digital converter, DAC: digital to analog converter, HPF: high-pass filter, LPF: low-pass filter, DDS: direct digital synthesizer, PS: phase shifter, EOPM: electro-optic phase modulator, EOAM: electro-optic amplitude modulator, AM: amplitude modulation, PI: proportional-integral control servo, PMT: photo-multiplier tube, MON: PC monitor.}
	\end{figure}
	
	The circuit implemented on the FPGA fabric via custom-made Verilog code can be divided into two main parts: the system for the atomic clock transition detection and locking of a cw laser to it (‘Transition Lock’), and the system for the RAM suppression (‘RAM Suppression’). Both are shown on Fig.~\ref{fig:FPGA_Fabric}. The transition lock is based on a direct digital synthesizer (DDS), which generates a sinusoidal modulation signal based on a 32-bit phase accumulator and a 16-bit lookup table stored in the FGPA-internal BRAM with additional Taylor interpolation to further reduce spurs. The DDS signal is outputted through one of the two RF-DACs and sent to an EO phase modulator to create the desired phase sidebands around the clock laser line. A second, synchronous DDS is used to produce a phase-shifted signal, which is then mixed with the fluorescence signal detected via a photo-multiplier tube and sampled through one of the two RF-ADCs. The signal from the photomultiplier tube is first fed through a digital high-pass filter to remove the DC and low-frequency flicker components before feeding it into the mixer. The high-pass filter is based on a finite-impulse response (FIR) lowpass filter, followed by a subtractor. This makes the filter immune against quantization noise. The output of the mixer is then fed through a configurable, 24-bit minimum-phase FIR low-pass filter to remove the high-frequency residuals and noise from the error signal. This signal runs through to two cascaded proportional-integral servos to stabilize the laser frequency to the atomic transition. The control signals are sent through two precision DACs for removing the fast and slow noise components to ensure stable long-term operation of the clock.
	
	The RAM suppression module detects the complex amplitude of the RAM signal, which is detected via a Si-PIN photodetector close to the Rb cell. The working principle for this is similar to a lock-in amplifier, which is synchronized to the modulation frequency of the transition lock: the RAM signal is sampled with the remaining RF-ADC, then down-sampled to increase the resolution, and band-pass filtered around the carrier frequency, followed by a pair of mixers driven with the sine and cosine waves from a third synchronous quadrature DDS. The global phase of this third DDS relative to the phase of the modulation DDS in the transition lock can be adjusted to account for the external cable delays and phase shifts in the EO phase modulator. The mixed signals are then again low-pass filtered with 24-bit resolution to generate the in-phase and quadrature components of the RAM signals with very high dynamic range. The values of these components are sent to a computer for monitoring and to aid in the initial RAM-servo fine-tuning process. The demodulated, complex RAM amplitudes are then used to amplitude modulate the sine and cosine waves, correspondingly, from a fourth quadrature DDS that is again phase adjusted relative to the main DDS to account for the cable and electronic delays between the FPGA board and the AM modulator. The amplitude modulated waves are amplified via adjustable-gain digital amplifiers, summed up, and outputted through the remaining fast DAC to drive the EO amplitude modulator.
	
	Overall, implementing this somewhat complex RF circuitry fully on the FPGA fabric ensures zero-drifts over long timescales and the absence of added flicker noise from the various mixers and amplifiers. The unavoidable quantization noise, on the other hand, can be made sufficiently small compared to the noise floor of the analog signals from the spectroscopic setup, such that the full available SNR can be maintained throughout the setup. One potential limitation is the maximum modulation frequency, which is currently limited to a few 10~MHz due to the 250~MHz ADC and DAC clock rates. However, this should be sufficient even for high-SNR applications, such as Pound-Drever-Hall locks of a cw-laser to a high-finesse cavity or for modulation-transfer in saturated absorption spectroscopy in atomic or molecular vapor clocks. The added benefit of a lower power consumption and autonomous locking and tuning options could play an important role for truly fieldable atomic clocks.
	
	\section{The two-photon rubidium clock}
\begin{figure}
		\centering\includegraphics[width = 0.5\textwidth]{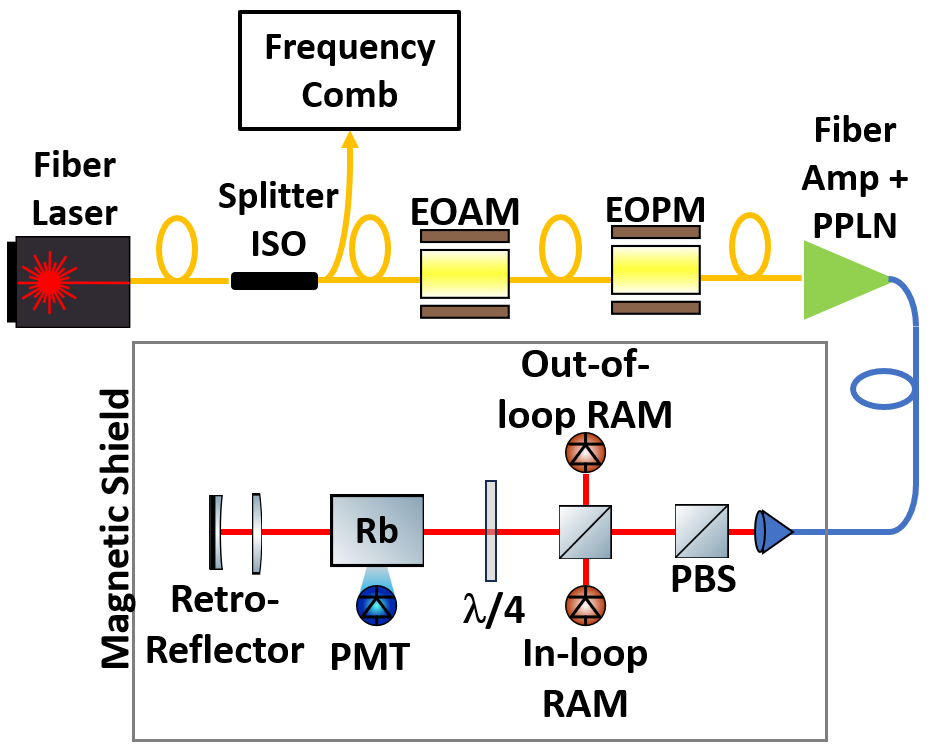}
		\caption{\label{fig:Rb2} FPGA circuits for generating error signal, locking the laser to the two-photon rubidium transition, and suppressing RAM. ADC: analog to digital converter, DAC: digital to analog converter, HPF: high-pass filter, LPF: low-pass filter, DDS: direct digital synthesizer, PS: phase shifter, EOPM: electro-optic phase modulator, EOAM: electro-optic amplitude modulator, AM: amplitude modulation, PI: proportional-integral control servo, PMT: photo-multiplier tube, MON: PC monitor.}
	\end{figure}

	The two-photon rubidium clock used to test the RAM suppression scheme relies on the 5S$_{1/2}\rightarrow$5D$_{5/2}$ (F = 2 $\rightarrow$ F' = 4) transition during which a rubidium atom absorbs two 778-nm photons to get excited to the 5D$_{5/2}$ level and then fluoresces a photon at 420~nm during the decay path. At the clock’s heart is an erbium fiber laser (Koheras ADJUSTIK, NKT Photonics) that lases at 1556~nm. The laser light is first split 5\% power for beating against a frequency comb using a combined isolator and splitter. The remaining 95\% power enters a zero-chirp EO amplitude modulator (21023819, JDSU), followed by an EO phase modulation (PM-OK5-10-PFA-PFA, EOSPACE) that is modulated at 200~kHz from a RF-DAC of the FPGA board. Both, the  EO amplitude modulator and the EO phase modulator have input and output ports of polarization-maintaining (PM1550) fibers. The zero-chirp nature of the EOAM ensures that the parasitic phase modulation is minimal during the application of amplitude modulation. The EO amplitude modulator is further biased at 2.1 V to ensure that most of the light goes to the output port. The modulated light then enters an erbium amplifier (Koheras BOOSTIK, NKT Photonics) and a periodically-poled lithium niobate (PPLN) waveguide (RSH-M0778-P15P85AL3, AdvR) to generate up to 50~mW at 778~nm. This frequency doubled light is then used to interrogate the atomic vapor, which is housed inside a temperature-controlled magnetic shield. After exiting the PM780 fiber inside the magnetic shield, the 778~nm light is first cleaned of residual unwanted polarization by a Glan-Laser polarizer, and sampled using a Si photodetector for both power stabilization and RAM suppression (`In-loop RAM' in Fig.~\ref{fig:Rb2}). The laser light is then turned into a circular polarization by means of a quarter waveplate before double-passing through the rubidium cell. A cat’s eye retro-reflector behind the rubidium cell ensures good beam overlap of the return pass. The reflected beam is sampled using a quadrature photodetector for beam pointing and RAM monitoring (`Out-of-loop RAM'). The fluorescence created by rubidium atoms when the laser frequency closely matches the two-photon transition is collected by a photo-multiplier tube (`PMT') positioned right next to the rubidium cell, which then fed to a RF-ADC of the FPGA via a transimpedance amplifier. To lock the fiber laser to a quarter of the frequency of the 5S$_{1/2}\rightarrow$5D$_{5/2}$ (F = 2 $\rightarrow$ F' = 4) transition, the FPGA board controls the voltage applied across a piezo element inside the laser, which effectively controls the laser frequency.  Once locked to the transition, the fluorescence emitted by the atoms is maximized. The rubidium cell is heated to 85~$^{\circ}$C from outside the magnetic shield via heat pipes to prevent stray magnetic field from entering the shield. However, due to the heat pipes’ large time delay from the heater to the rubidium cell, a fiber-coupled and wavelength-locked 980-nm diode laser with 100~mW output power is used to fine-tune the temperature of the cold finger of the rubidium cell at short-time scales. 
	
	The frequency of the clock was compared against a frequency comb that had its carrier-offset frequency stabilized and was locked to an ultra-stable optical cavity for short-term stability. The cavity drift over time manifested as drifts in the comb’s repetition rate and was monitored against a hydrogen maser that was disciplined to a cesium fountain clock. Both the comb’s repetition rate and the frequency of the beat note between the clock laser and the frequency comb were recorded using a home-built, FPGA-based, dead-time-free, and synchronous multi-channel frequency counter. However, since the repetition rate of the comb was at 250~MHz, which was the sampling rate of the frequency counter, the repetition rate of the comb was mixed down to 13.5~MHz using a synthesizer that was phase-locked to the maser signal. The clock’s long-term stability was obtained by subtracting the drift of the comb’s measured repetition rate from the beat note between the clock and the comb.
	
	\section{Results}
\begin{figure}
		\centering\includegraphics[width = 0.495\textwidth]{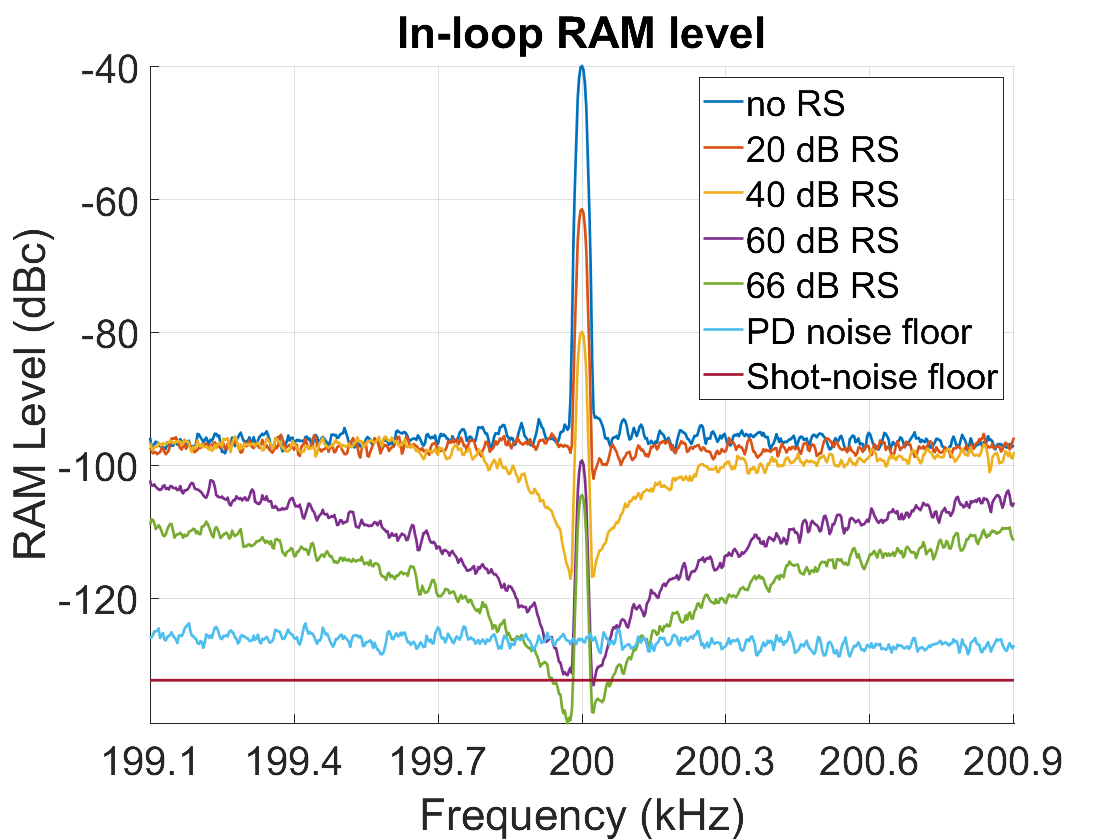}
		\includegraphics[width = 0.495\textwidth]{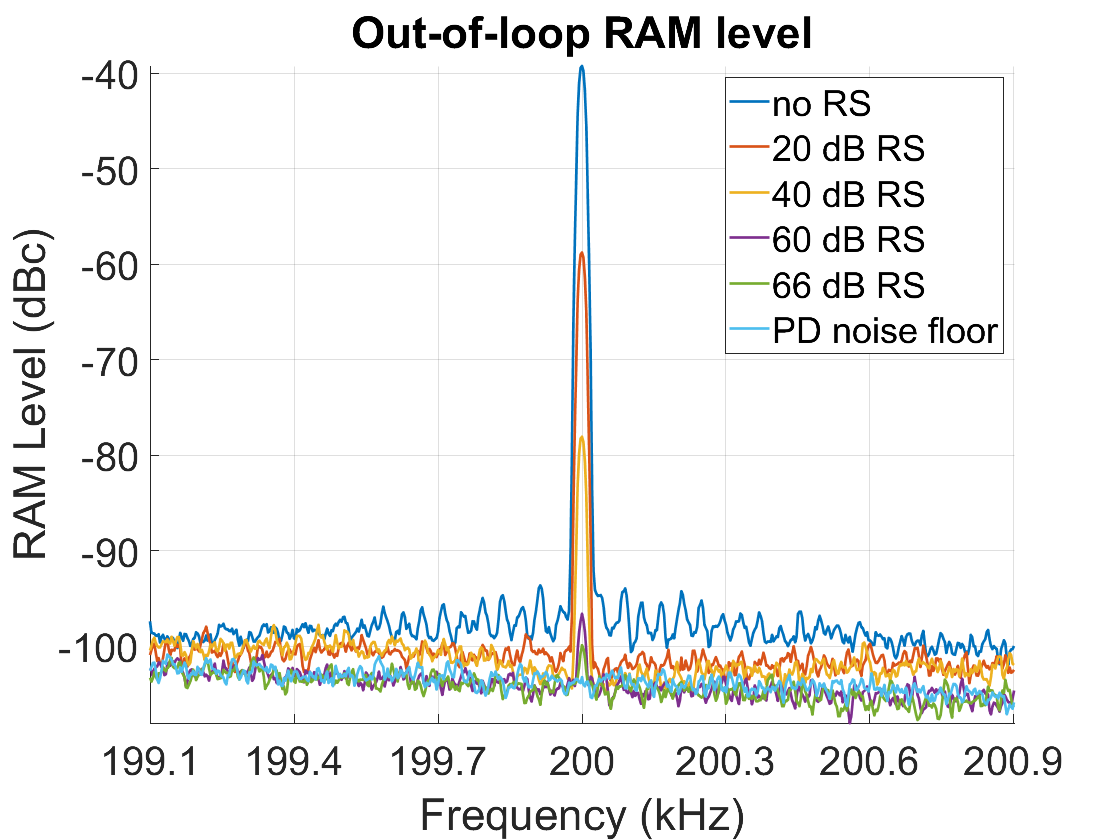}
		\caption{\label{fig:RAM_inloop_mon}The performance of RAM suppression observed via (left) in-loop and (right) out-of-loop detectors for no RAM suppression (RS), 20 dB RS, 40 dB RS, 60 dB RS, 66 dB RS, and their noise floor and the calculated shot noise floor (for (left) only) normalized against the respective carrier powers. The RBW was 10~Hz and the number of averages 20 for all plots.}
	\end{figure}

	To demonstrate the performance of the RAM suppression scheme, we measured the RAM levels at both, in-loop and out-of-loop RAM monitors (see Fig.~\ref{fig:Rb2}) using an RF-spectrum analyzer. The measurement runs were performed for varying levels of RAM suppression. The measured RAM levels were normalized against the respective carrier powers on each photodiode (PD) to obtain the RAM values in units of dBc (dB relative to the carrier). Different RAM suppression (RS) levels were applied by changing the gain of the adjustable-gain amplifiers in the FPGA fabric. As shown in Fig.~\ref{fig:RAM_inloop_mon}, the RAM levels decrease with increasing RS levels, up to 64~dB for the in-loop signal, and 60~dB for out-of-loop signal. The maximum achievable RAM suppression was limited firstly by the SNR of the two detectors used and then by the shot-noise limit. The noise of the in-loop detector at -126 dBc in 10~Hz and the calculated shot-noise floor at -132~dBc in 10~Hz limited the allowable maximum RAM servo gain as not to add detector and shot noises to the laser relative intensity noise (RIN) (Fig.~\ref{fig:RAM_inloop_mon}, left; the RIN suppression starts to dip below the detector noise floor for gain of 60 and the calculated shot-noise floor for gain of 66~dB). While the RAM suppression might still improve with higher gains, the added RIN noise might start to affect the clock’s stability. The out of loop measurement further limited the assessment of the achieved RAM suppression due to the rather high noise floor of the out-of-loop detector (-104~dBc over 10~Hz) compared to the calculated shot-noise floor at -138~dBc over 10~Hz. That detector is based on a large area quadrant diode, which was initially intended for beam-pointing monitoring and control. The large capacitance of that diode, paired with the large voltage noise of the transimpedance amplifier significantly limited the performance of that detector. Therefore, the demonstrated 60~dB of out-of-loop RAM suppression (corresponding to a RAM level of -100~dBc over 10~Hz bandwidth) is likely an upper bound. The high noise floor of the out-of-loop detector also prevented us from observing the feared increase of laser RIN in the vicinity of the modulation frequency for large RAM servo gains. Better detectors and higher optical power might well allow further suppression of RAM to below -100~dBc residuals. However, we should note that such low levels of RAM might bring no further benefit to the clock’s stability as discussed below. 
	
\begin{figure}
		\centering\includegraphics[width = 0.495\textwidth]{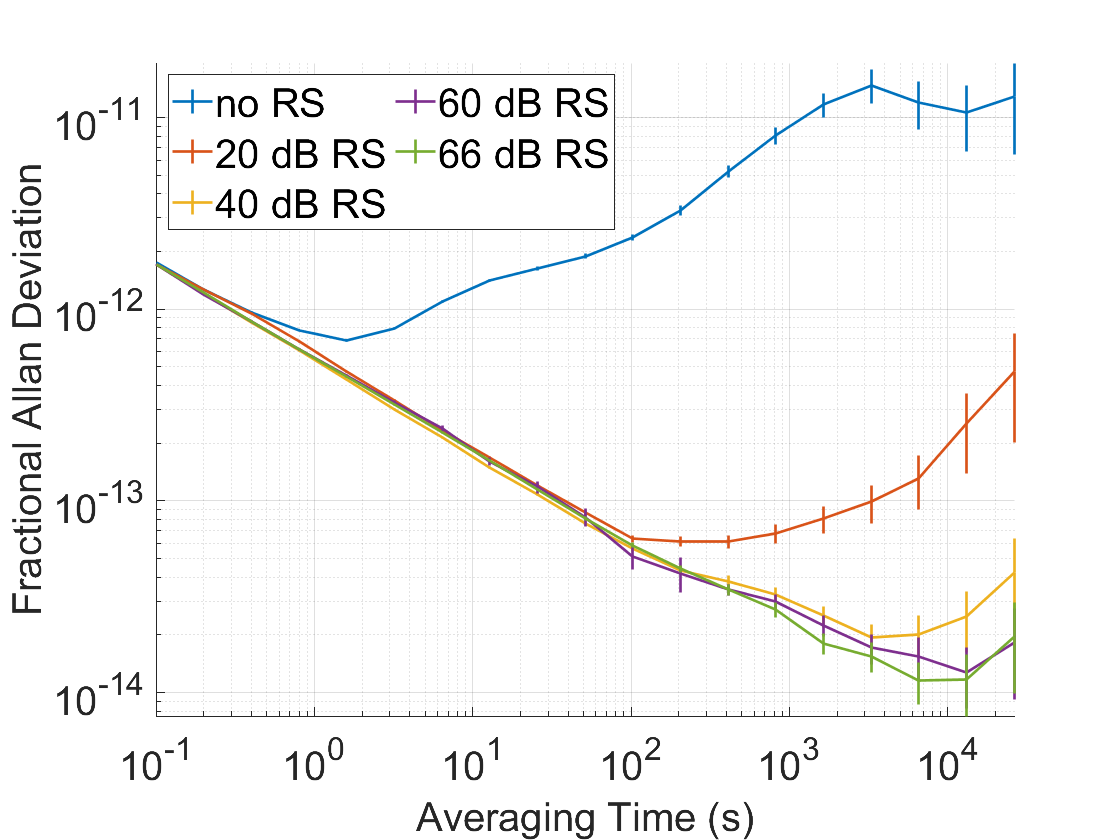}
		\caption{\label{fig:Allan_plot} The fractional instability of the clock for different levels of RAM suppressions.}
	\end{figure}

	The measured fractional instability of the two-photon rubidium frequency standard for different levels of RAM suppressions are shown in Fig.~\ref{fig:Allan_plot}. The frequency data for `no RS', and `20~dB RS' were recorded over 17 hours, while the frequency data for the remaining RAM suppression levels were recorded over 28~hours each. For averaging times of 200~s and below, the fractional instability of the clock was obtained directly against the ultra-stable cavity. For averaging times above 200~s, the measured frequency drift of the cavity was subtracted from the frequency data before calculating the fractional Allan deviations. The fractional instability of the Rb clock starts at $2\times10^{-12}$ at 0.1~s then decreases as $1/\sqrt{\tau}$ except when there is no RAM suppression. Without RAM suppression, the stability already deteriorates at 1~s and increases to more than $10^{-11}$ at 1000~s. When the FPGA board was configured to reduce the RAM level by 20~dB, the fractional instability of the clock continues to decrease down to $6\times10^{-14}$ at 100~s but starts to suffer from the remaining RAM after that. For `40~dB RS', the fractional instability of the clock continued to improve to $2\times10^{-14}$ at 3000~s averaging time before again being limited by the remaining RAM. The fractional instabilities for `60~dB RS' and `66~dB RS' were almost identical, albeit lower than that of `40~dB RS', indicating that the clock’s performance was likely no longer limited by the remaining RAM, but rather other instabilities in the clock setup. The increase in fractional instabilities for these two suppression levels at averaging time of $10^{4}$~s and longer could for instance be due to thermal drifts or laser power fluctuations.
	
	\section{Discussion}
	It is worth noting that a RAM suppression of 20~dB resulted in a stability improvement by a factor of more than 10 for averaging times longer than 10~s. This was against our expectation that a 20-dB reduction in RAM should only result in a factor of 10 reduction in the fractional instability and we speculate that this might be due to the large, unbound frequency random walks and drifts with no RAM servo active.
	
	An interesting aspect is that this complex modulation scheme can also be applied when higher modulation frequencies are needed, such as for locking a cw-laser to an ultra-stable cavity, where a modulation frequency of the order of 10~MHz is often desirable. The processing delay of the FPGA-based RAM suppression algorithm should indeed not affect the performance even at 10s of MHz modulation frequencies, as the digital phase adjustments allow the cancellations of both, the cable and processing delays. As seen in Fig.~\ref{fig:Allan_plot}, the effect of RAM on the fractional instability only becomes significant for long-enough averaging times (e.g., for > 0.5 s in the case of the Rb clock). This means that the amplitude of the feedback signal to the EO amplitude modulator needs to be corrected only over relatively slow timescales and the carrier frequency itself is not important as long as the phase of the feedback signal relative to the modulation signal remains unchanged. We found that it was the case for over half a year during which this suppression scheme was implemented even when neither the temperature of the EO phase modulator nor the bias voltage to the EO amplitude modulator were stabilized. In other words, the phase delay between the modulation signal and the RAM suppression signal due to the fiber, cable and processing delays are indeed time independent; this is in stark contrast to the phase stability of analog phase shifters and analog filters.
	
	Although the RAM suppression scheme should be readily applicable to locking a cw-laser to an ultra-stable cavity, the feedback through the slow ADCs might not be fast enough for low-noise applications, such as photonic microwave generation or pre-stabilization of laser to sub-Hz linewidth. In such case, the feedback can be carried out through the modulation output port similar to a method by Endo et al.\cite{Endo2018} However, the time delays of the ADCs and DACs limit the loop bandwidth to 1~MHz leading to a higher phase noise compared to what could be achieved with an analog loop filter. Despite all that, we believe that phase noise level should be many times lower than by just feeding back to the fiber laser’s PZT port. 
	
	Last but not least, we believe that the limit of our RAM suppression setup in the demonstration with a two-photon rubidium clock was due to the high noise floors of the in-loop and out-of-loop detectors. This could be overcome with well-designed photodetectors that have lower noise floors than the shot noise floor of the sampled laser beam. This could potentially result in a RAM level close to the shot noise limit. Even with the current limitations, the remaining RAM level of -100~dBc, or equivalently 10 ppm in amplitude, is comparable to those of the best passive or active RAM suppression techniques mentioned above. 
	
	\section{Conclusion}
	We developed an FPGA fabric that houses the entire complex modulation method for RAM suppression and error signal generation for locking lasers to atomic or cavity frequency references with all of the servos on a low-cost, off-the-shelf FPGA chip that is free of drifts and flicker noise. We demonstrated that the stability of a two-photon rubidium clock can be improved by three decades with the remaining RAM level at -100~dBc, or equivalently 10~ppm that and that RAM is no longer the limiting factor of the clock stability We found that this level of RAM suppression could be maintained over half a year of testing with no signs of degradation. Since this suppression method does not have any specific requirements for the EO phase modulator such as temperature control, DC port, wedge crystal, or special crystal composition, this scheme can be applied to almost all of the EO phase modulators available on the market for a price of somewhat larger optical insertion loss. Together with fiber-based EO phase and amplitude modulators, this technique could enable compact, truly fieldable, and high-performing frequency standards.
	
	\section*{Acknowledgement}
	The development of the atomic clock’s physics package was in part supported by the Air Force Research Grant No. AWD-20-10-0237. The authors would like to thank the National Institute of Standard and Technology, Matthew Hummon, and John Kitching’s group for lending space in their labs and the use of their frequency comb stabilized to an ultra-stable cavity and providing access to a Cs-disciplined hydrogen maser.
	
	\section*{Author declarations}
	\subsection*{Conflict of Interest}
		The authors have no conflicts of interest to disclose.
	\subsection*{Author Contributions}
		All authors contributed equally to the research and the writing of this manuscript.
	\section*{Data availability}
		The data that support the findings of this study are available from the corresponding author upon reasonable request.
	\bibliography{references}
		
	\end{document}